\documentclass[11pt]{article}
\usepackage{amsfonts}

\usepackage{amssymb}
\usepackage{latexsym}
\usepackage{graphicx}
\usepackage[english]{babel}

\begin{document}
\input epsf

\makeatletter
\@addtoreset{equation}{section}
\makeatother


\begin{flushright}
\end{flushright}
\vspace{20mm}

 \begin{center}
{\LARGE Excitations in the deformed D1D5 CFT}
\\
\vspace{18mm}
{\bf   Steven G. Avery$^\Diamond$\footnote{avery@mps.ohio-state.edu}, Borun D. Chowdhury$^\sharp$\footnote{b.d.chowdhury@uva.nl} and Samir D. Mathur$^\Diamond$\footnote{mathur@mps.ohio-state.edu}
\\}
\vspace{8mm}
$\Diamond$ 
Department of Physics,\\ The Ohio State University,\\ Columbus,
OH 43210, USA\\ 
\vspace{8mm}
$\sharp$
Instituut voor Theoretische Fysica,\\
Universiteit van Amsterdam,\\
Amsterdam 1018 XE, The Netherlands\\
\end{center}
\vspace{10mm}

\thispagestyle{empty}
\begin{abstract}

We perform some simple computations for the first order deformation of the D1D5 CFT off its orbifold point. It had been shown earlier that under this deformation the vacuum state changes to a squeezed state (with the further action of a supercharge). We now start with states containing one or two initial quanta and write down the corresponding states obtained under the action of deformation operator. The result is relevant to the evolution of an initial excitation in the CFT dual to the near extremal D1D5 black hole:  when a left and a right moving excitation collide in the CFT, the deformation operator spreads their energy over a larger number of quanta, thus evolving the state towards the infrared.

\end{abstract}
\newpage
\renewcommand{\theequation}{\arabic{section}.\arabic{equation}}

\def\nn{\nonumber \\}
\def\p{\partial}
\def\h{{1\over 2}}
\def\be{\begin{equation}}
\def\bea{\begin{eqnarray}}
\def\ee{\end{equation}}
\def\eea{\end{eqnarray}}
\def\r{\rightarrow}
\def\tildr{\tilde}
\def\n{\nonumber}
\def\nn{\nonumber \\}
\def\t{\tilde}
\def\b{\bigskip}
\newcommand{\Nsc}{\mathcal{N}}
\newcommand{\bj}{\bar{\jmath}}
\def\sqi{{1\over \sqrt{2}}}
\newcommand\eqref[1]{(\ref{#1})}

\newcommand\ket[1]{|#1\rangle}
\newcommand\bra[1]{\langle #1|}
\newcommand\com[2]{[#1,\,#2]}
\newcommand\ac[2]{\{#1,\,#2\}}

\section{Introduction}
\label{intr}\setcounter{equation}{0}

The D1D5 bound state provides a simple system to study the physics of black holes\cite{counting, radiation, fuzzballs, adscft}. It is believed that at a particular value of couplings the CFT reaches an `orbifold point' where it is a 1+1 dimensional sigma model with target space an orbifold space. The CFT at the orbifold point is comparativey simple to analyze, and several agreements are observed between the thermodynamic properties of this CFT and the physics of black holes. But the physics of black holes does not correspond to the orbifold point, and we have to deform away from this special point in the CFT moduli space to fully analyze all the aspects of the gravitational physics \cite{deformation}. 

In \cite{acm1} the effect of the deformation operator was computed on the simplest state of the CFT: the lowest spin Ramond vacuum of the `untwisted' sector. It was found that the resulting state could be written in closed form. The essential structure of this state that of a squeezed state: we have an exponential of a bilinear in bosonic and fermionic creation operators (schematically, $ e^{\gamma^B \alpha^\dagger \alpha^\dagger+ \gamma^F d^\dagger d^\dagger}$). The full state is given by the action of a supercharge $G$ on this squeezed state.

To understand the evolution of states in the deformed CFT we have to consider initial states that have one or more initial excitations. Suppose a left and a right moving excitation come near each other in the CFT. The deformation operator can take the energy and momentum of this pair and distribute it over a different set of excitations. In particular the exponential term $ e^{\gamma^B \alpha^\dagger \alpha^\dagger+ \gamma^F d^\dagger d^\dagger}$ is always created by the deformation operator, and we see that this can lead to an arbitrary number of excitations sharing the energy and momentum of the initial colliding pair.

\begin{figure}[ht]
\begin{center}
\includegraphics[width=4cm]{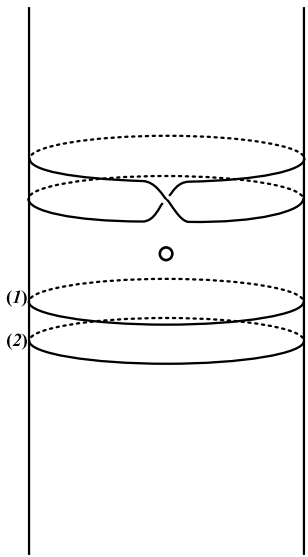}
\caption{The effect of the twist contained in the deformation operator: two circles at earlier times get joined into one circle after the insertion of the twist.}\label{2-twist}
\end{center}
\end{figure}

In this paper we will consider the action of the deformation operator on states that contain bosonic and fermionic excitations in the initial state. The deformation operator $\hat O$ is made by the action of a supercharge $G^-$ on a twist $\sigma_2^+$ (shown in figure \ref{2-twist}). We first work out the effect of $\sigma_2^+$ on 
an initial state containing a single boson or  a single femion.   We then work out  cases where two bosonic operators or two fermionic operators are present in the initial state. Apart from the term where each initial mode becomes a set of modes in the final state, we also get a contribution from a `Wick contraction'  between the two modes in the initial state. Finally we take the case where a single boson is present in the initial state, and work out the complete final state that results, now including the effect of the supercharge $G^-$. These computations illustrate a set of methods that can be used to find the effect of the deformation operator on any initial state.

\section{The D1D5 CFT at the orbifold point}\label{sectiontwo}

In this section we summarize some properties of the D1D5 CFT at the orbifold point and the deformation operator that we will use to perturb away from the orbifold point. For more details, see \cite{acm1}.

\subsection{The CFT}

Consider type IIB string theory, compactified as
\be
M_{9,1}\rightarrow M_{4,1}\times S^1\times T^4.
\label{compact}
\ee
Wrap $N_1$ D1 branes on $S^1$, and $N_5$ D5 branes on $S^1\times
T^4$. The bound state of these branes is described by a field
theory. We think of the $S^1$ as being large compared to the $T^4$, so
that at low energies we look for excitations only in the direction
$S^1$.  This low energy limit gives a conformal field theory (CFT) on
the circle $S^1$.

We can vary the moduli of string theory (the string coupling $g$, the
shape and size of the torus, the values of flat connections for gauge
fields etc.). These changes move us to different points in the moduli
space of the CFT. It has been conjectured that we can move to a point
called the `orbifold point' where the CFT is particularly simple
\cite{orbifold}. At this orbifold point the CFT is
a 1+1 dimensional sigma model. We will work in the Euclidized theory, where
the base space is a cylinder spanned by the coordinates 
\be
\tau, \sigma: ~~~0\le \sigma<2\pi, ~~~-\infty<\tau<\infty
\ee
The target space of the sigma model is the `symmetrized product' of
$N_1N_5$ copies of $T^4$,
\be
(T_4)^{N_1N_5}/S_{N_1N_5},
\ee
with each copy of $T^4$ giving 4 bosonic excitations $X^1, X^2, X^3,
X^4$. It also gives 4 fermionic excitations, which we call $\psi^1,
\psi^2, \psi^3, \psi^4$ for the left movers, and $\bar\psi^1,
\bar\psi^2,\bar\psi^3,\bar\psi^4$ for the right movers. The fermions can be
antiperiodic or periodic around the $\sigma$ circle. If they are
antiperiodic on the $S^1$ we are in the Neveu-Schwarz (NS) sector, and
if they are periodic on the $S^1$ we are in the Ramond (R)
sector. The central charge of the theory with fields
$X^i, \psi^i, ~i=1\dots 4$ is
\be
c=6
\ee
The total central charge of the entire system is thus $6 N_1N_5$.

\subsection{Symmetries of the CFT}

The D1D5 CFT has $(4,4)$ supersymmetry, which means that we have
$\mathcal{N}=4$ supersymmetry in both the left and right moving
sectors. This leads to a superconformal ${\cal N}=4$ symmetry in both
the left and right sectors, generated by operators $L_{n}, G^\pm_{r},
J^a_n$ for the left movers and $\bar L_{n}, \bar G^\pm_{r}, \bar
J^a_n$ for the right movers. The algebra generators and their OPEs and commutators are given in Appendix~\ref{ap:CFT-notation}. 

Each $\Nsc = 4$ algebra has an internal R symmetry group
SU(2), so there is
a global symmetry group $SU(2)_L\times SU(2)_R$.  We denote the
quantum numbers in these two $SU(2)$ groups as
\be
SU(2)_L: ~(j, m);~~~~~~~SU(2)_R: ~ (\bj, \bar m).
\ee
In the geometrical setting of the CFT, this symmetry arises from the
rotational symmetry in the 4 space directions of $M_{4,1}$ in
Equation~\eqref{compact},
\be
SO(4)_E\simeq SU(2)_L\times SU(2)_R.
\label{pthree}
\ee
Here the subscript $E$ stands for `external', which denotes that these
rotations are in the noncompact directions. These quantum numbers
therefore give the angular momentum of quanta in the gravity
description.  We have another $SO(4)$ symmetry in the four directions
of the $T^4$. This symmetry we call $SO(4)_I$ (where $I$ stands for
`internal'). This symmetry is broken by the compactification of the
torus, but at the orbifold point it still provides a useful organizing
principle. We write
\be
SO(4)_I\simeq SU(2)_1\times SU(2)_2.
\ee
We use spinor indices $\alpha, \dot\alpha$ for $SU(2)_L$ and $SU(2)_R$
respectively. We use spinor indices $A, \dot A$ for $SU(2)_1$ and
$SU(2)_2$ respectively.

The 4 real fermions of the left sector can be grouped into complex
fermions $\psi^{\alpha A}$. The right fermions have indices $\bar{\psi}^{\dot\alpha \dot A}$. The bosons $X^i$ are a vector in the
$T^4$. They have no charge under $SU(2)_L$ or $SU(2)_R$ and are
given by
\begin{equation}
[X]_{\dot{A}A} = {1\over \sqrt{2}}X^i(\sigma^i)_{\dot{A}A}.
\end{equation}
where $\sigma^i, i=1, \dots 4$ are the three Pauli matrices and the identity. (The
notations described here are explained in full detail in
Appendix~\ref{ap:CFT-notation}.)

\subsection{States and operators}

Since we orbifold by the symmetric group $S_{N_1N_5}$, we generate
`twist sectors', which can be obtained by acting with `twist
operators' $\sigma_n$ on an untwisted state. Each set of linked copies will be termed a `component string'. 
The simplest states of the CFT are in the `untwisted sector' where no copy of the $c=6$ CFT is linked to any other copy; i.e. all component strings have winding number unity.  Consider one component string, and consider the theory defined on the cylinder.  The fermions on this string can be either periodic around the $\sigma$ circle of the cylinder (Ramond sector R) or antiperiodic (Neveu-Schwarz sector NS). Consider one copy of the $c=6$ CFT. The simplest state of this theory is the NS sector vacuum 
\be
|0\rangle_{NS}: ~~~h=0, ~~m=0
\label{nsvac}
\ee
But the physical  CFT arising from the D1D5 brane bound state is in the Ramond (R)
sector.  We can relate the state 
(\ref{nsvac}) to a Ramond ground state using spectral flow \cite{spectralref}. Spectral flow maps amplitudes in the CFT to amplitudes in another CFT; under this map  dimensions and charges change as (we write only the left sector)
\be
h'=h+\alpha j +{c\alpha^2\over 24}, ~~~
j'=j+{c\alpha\over 12}
\label{spectral}
\ee
We have $c=6$. Setting $\alpha=-1$ gives
\be
|0^-_R\rangle: ~~h={1\over 4}, ~~~m=-\h
\ee
which is one of the Ramond ground states of the $c=6$ CFT for a component string with winding number unity. Other Ramond ground states are obtained by acting with fermion zero modes, so that we have four states in all
\be
|0_R^-\rangle,~~~\psi^{++}_0|0_R^-\rangle, ~~~\psi^{+-}_0|0_R^-\rangle, ~~~\psi^{++}_0\psi^{+-}_0|0_R^-\rangle
\label{rground}
\ee
(with similar possibilities for the right moving sector).

The deformation operator involves the twist $\sigma_2$. As we go around a point of insertion of this twist, the fermions in the first copy change to fermions in the second copy, and after another circle return to their original value. Creating such a twist automatically brings in a `spin field' at the insertion point, which has $h={1\over 2}, j=\h$ \cite{lm1,lm2}. Thus there are two possible insertions of such a twist, with $m=\h$ and with $m=-\h$. We write these as
$\sigma_2^+$ and $\sigma_2^-$
respectively. The operator $\sigma_2^+$ is a chiral primary and $\sigma_2^-$ is an anti-chiral primary.

\subsection{The deformation operator}

The deformation operator is a singlet under $SU(2)_L\times SU(2)_R$.
To obtain such a singlet we apply modes of $G^\mp_{\dot A}$ to
$\sigma_2^\pm$. In \cite{acm1} it was shown that we can write the deformation operator as
\be
\hat O_{\dot A\dot B}(w_0)=\Big [{1\over 2\pi i} \int _{w_0} dw G^-_{\dot A} (w)\Big ]\Big [{1\over 2\pi i} \int _{\bar w_0} d\bar w \bar G^-_{\dot B} (\bar w)\Big ]\sigma_2^{++}(w_0)
\ee
The left and right movers separate out completely for all the computations that we will perform. Thus from now on we will work with the left movers only; in particular the twist operator will be written only with it left spin: $\sigma_2^+$.

The operator $\sigma^+_2$ is normalzied to have a unit OPE with its conjugate
\be
\sigma_{2,+}(z')\sigma_2^{+}(z)\sim {1\over (z'-z)}
\ee
Acting on the Ramond vacuum this implies \cite{acm1}
\be
\sigma_2^+(w)|0_R^-\rangle^{(1)}\otimes |0_R^-\rangle^{(2)}=|0_R^-\rangle+\dots
\label{qwthree}
\ee
Here $|0_R^-\rangle$ is the spin down Ramond vacuum of the CFT on the doubly wound circle produced after the twist. The normalization (\ref{qwthree}) has given us the coefficient unity for the first term on the RHS, and the `$\dots$' represent excited states of the CFT on the doubly wound circle.

\section{The effect of the deformation operator on the Ramond vacuum}

In this section we summarize the results of \cite{acm1} which computed the effect of  the deformation operator acts on a very simple state: the Ramond vacuum of the lowest spin state in the untwisted sector Ramond sector.

\begin{figure}[ht]
\begin{center}
\includegraphics[width=4cm]{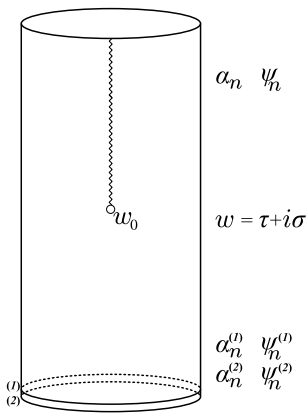}
\caption{Before the twist insertion we have boson and fermion modes on two copies of the $c=6$ CFT. These modes are labeled with superscripts $(1), (2)$ respectively. The twist inserted at $w_0$ joins these to one copy for $\tau>\tau_0$; the modes here do not carry a superscript. The branch cut above $w_0$ indicates that we have two sets of fields at any given $\sigma$; these two sets go smoothly into each other as we go around the cylinder, giving a continuous field on a doubly wound circle.}\label{two}
\end{center}
\end{figure}

We consider the amplitude depicted in figure~\ref{two}.

In the initial state we have two copies of the $c=6$ CFT, living on `single circles' with $0\le \sigma<2\pi$. We take the initial state
\be
|\psi\rangle_i=|0_R^{-}\rangle^{(1)}\otimes |0_R^{-}\rangle^{(2)}
\label{initial}
\ee

We insert the operator $\hat O_{\dot A\dot B}$ at the point $w_0$ on the cylinder.  Thus the final state that we want to find is given by
\be
|\psi\rangle_f=\hat O_{\dot A}(w_0)|\psi\rangle_i=\Big [{1\over 2\pi i}\int_{w_0} dw G^-_{\dot A}(w)\Big ]~\sigma_2^{+}(w_0)
|0_R^{-}\rangle^{(1)}\otimes |0_R^{-}\rangle^{(2)}
\label{pone}
\ee
The final state will contain one component string with winding number $2$, since the deformation operator contains the twist $\sigma_2^+$.

\subsection{Mode expansions on the cylinder}\label{expansions}

We expand the bosonic and fermionic field in modes on the cylinder.  Below the twist insertion ($\tau<\tau_0$) we have
\bea
\alpha^{(i)}_{A\dot A, n}&=& {1\over 2\pi} \int_{\sigma=0}^{2\pi} \p_w X^{(i)}_{A\dot A}(w) e^{nw} dw, ~~~i=1,2
\label{lone}\\
d^{(i)\alpha A}_n&=&{1\over 2\pi i} \int_{\sigma=0}^{2\pi} \psi^{(i)\alpha A}(w) e^{nw} dw, ~~~i=1,2
\label{loneqq}
\eea
The commutation relations are
\bea
[\alpha^{(i)}_{A\dot A, m}, \alpha^{(j)}_{B\dot B, n}]&=&-\epsilon_{AB}\epsilon_{\dot A\dot B}\delta^{ij} m \delta_{m+n,0}\\
\{ d^{(i)\alpha A}_m, d^{(j)\beta B}_n\}&=&-\epsilon^{\alpha\beta}\epsilon^{AB}\delta^{ij}\delta_{m+n,0}
\eea

Above the twist insertion ($\tau>\tau_0$)  we have a doubly twisted circle. The modes are\bea
\alpha_{A\dot A, n}&=& {1\over 2\pi} \int_{\sigma=0}^{4\pi} \p_w X_{A\dot A}(w) e^{{n\over 2}w} dw
\label{qaone}\\
d^{\alpha A}_n&=& {1\over 2\pi i}\int_{\sigma=0}^{4\pi} \psi^{\alpha A}(w) e^{{n\over 2} w} dw
\label{qaoneqq}
\eea
The commutation relations turn out to be
\bea\label{bcommtwist}
[\alpha_{A\dot A, m}, \alpha_{B\dot B, n}]&=&-\epsilon_{AB}\epsilon_{\dot A\dot B} m \delta_{m+n,0}\\
\label{fcommtwist}
\{ d^{\alpha A}_m, d^{\beta B}_n\}&=&-2\epsilon^{\alpha\beta}\epsilon^{AB}\delta_{m+n,0}
\eea

\subsection{The supercharge}\label{scharge}

\begin{figure}[ht]
\begin{center}
\includegraphics[width=8cm]{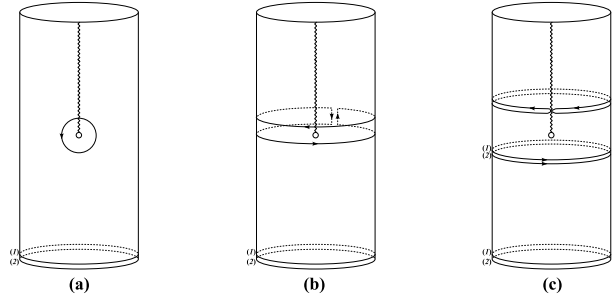}
\caption{(a) The supercharge in the deformation operator is given by integrating $G^-_{\dot A}$ around the insertion at $w_0$. (b) We can stretch this contour as shown, so that we get a part above the insertion and a part below, joined by vertical segments where the contributions cancel. (c) The part above the insertion gives the zero mode of the supercharge on the doubly wound circle, while the parts below give the sum of this zero mode  for each of the two initial copies of the CFT.}\label{G-contour}
\end{center}
\end{figure}

Let us first put the ${1\over 2\pi i}\int_{w_0} dw G^-_{\dot A}(w)$ operator in (\ref{pone}) in a more convenient form. The contour in this operator runs circles the insertion $w_0$ (figure~\ref{G-contour}(a)). We  can stretch this to a contour that runs around the rectangle shown in figure~\ref{G-contour}(b). The vertical sides of the contour cancel out.  
We can thus break the contour into a part above the insertion and a part below the insertion (fig.~\ref{G-contour}(c)). The lower leg gives 
\be
-{1\over 2\pi i} \int_{w=\tau_0-\epsilon}^{\tau_0-\epsilon+2\pi i}G^-_{\dot A}(w)=-{1\over 2\pi i} \int_{w=\tau_0-\epsilon}^{\tau_0-\epsilon+2\pi i}[G^{(1)-}_{\dot A}(w)+G^{(2)-}_{\dot A}(w)]dw \equiv -\Big (G^{(1)-}_{\dot A, 0}+G^{(2)-}_{\dot A, 0}\Big )
\label{ltwo}
\ee
 The upper leg gives
\be
{1\over 2\pi i} \int_{w=\tau_0+\epsilon}^{\tau_0+\epsilon+4\pi i}G^-_{\dot A}(w)dw\equiv G^-_{\dot A, 0}
\label{lfive}
\ee
where we note that the two copies of the CFT have linked into one  copy on a doubly wound circle, and we just get the zero mode of $G^-_{\dot A}$ on this single copy.

Note that
\be
G^{(i)-}_{\dot A, 0}|0_R^{--}\rangle^{(i)}=0, ~~~i=1,2
\label{lthree}
\ee
since the $\alpha$ index of $|0_R^{\alpha -}\rangle^{(i)}$ forms a doublet under $SU(2)_L$, and 
we cannot further lower the spin of $|0_R^{--}\rangle^{(i)}$ without increasing the energy level. Thus when the initial state (\ref{initial}), the lower contour gives nothing, and we have
\be
|\psi\rangle=G^-_{\dot A, 0}\sigma_2^{+}(w_0)
|0_R^{-}\rangle^{(1)}\otimes |0_R^{-}\rangle^{(2)}
\label{ptwo}
\ee
Let us write this as 
\be
|\psi\rangle=G^-_{\dot A, 0}|\chi\rangle
\ee
where
\be
|\chi\rangle=\sigma_2^{+}(w_0)
|0_R^{-}\rangle^{(1)}\otimes |0_R^{-}\rangle^{(2)}
\label{chiq}
\ee

\subsection{Steps in finding $|\chi\rangle$}\label{steps}

We start from the original problem, which is in the Ramond sector on the cylinder, and make a sequence of spectral flows and coordinate changes to map it to  a simpler form. (For more details, see \cite{acm1}.)

\b

(a) First we perform a spectral flow (\ref{spectral}) with parameter $\alpha=1$. This brings the two copies of the $c=6$ CFT in the initial state to the NS sector.

\b

(b) We wish to go to a covering space which will allow us to see explicitly the action of the twist operator. First map the cylinder with coordinate $w$ to the plane with coordinate $z$
\be
z=e^w
\ee

\begin{figure}[ht]
\begin{center}
\includegraphics[width=5cm]{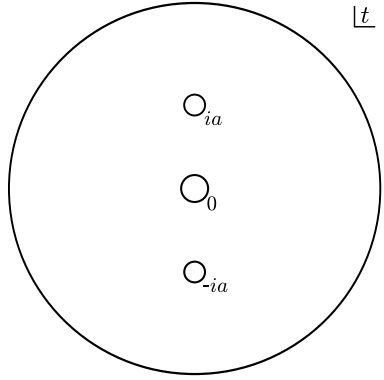}
\caption{The $z$ plane is mapped to the covering space -- the $t$ plane -- by the map $z=z_0+t^2$. The point $z=0$ corresponds to $\tau\r-\infty$ on the cylinder, and the two copies of the CFT there correspond to the points $t=\pm ia$. The location of the twist operator maps to $t=0$. The top the cylinder $\tau\r\infty$ maps to $t\r \infty$. After all maps and spectral flows, we have the NS vacuum at $t=0, \pm ia$, and so we can smoothly close all these punctures. The state $|\chi\rangle$ is thus just the $t$ plane vacuum; we must write this in terms of the original cylinder modes and apply the supercharge to get the final state $|\psi\rangle$.}\label{t-plane}
\end{center}
\end{figure}

We then pass to the cover of the $z$ plane via the map
\be
z=z_0+t^2; ~~~~~z_0=e^{w_0}\equiv a^2
\ee
Since step (a) mapped the states at $\tau\r-\infty$ to the NS vacuum, we have just the NS vacuum at the corresponding punctures $t=\pm ia$ in the $t$ plane, and we can smoothly close these punctures. 
The location of the twist insertion $\sigma_2^+$ maps to $t=0$. At this location we have the state $|0^+_R\rangle_t$, the spin up Ramond vacuum of the $t$ plane (figure \ref{t-plane}).

\b

(c) We  perform a spectral flow with $\alpha=-1$ in the $t$ plane.  The Ramond ground state at $t=0$ maps to the NS vacuum in the $t$ plane
\be
|0^+_R\rangle_t\r |0\rangle_t
\ee
We can now smoothly close the puncture at $t=0$ as well. 

\b

Thus we see that under this set of maps the state (\ref{initial}) just generates a state which is the $t$ plane vacuum at large $t$: the punctures at $t=\pm ia$ and at $t=0$ have all been smoothly closed with no insertions. Since $t$ large corresponds to $\tau\r\infty$ we see that the state $|\chi\rangle$ in the present description is just the $t$ plane vacuum. But though we have now found the state created by the action of $\sigma_2^+$, we still  need to express this state in terms of the modes appropriate to the original problem; i.e., the modes on the cylinder before any spectral flows or coordinate changes. 

Thus we need to know how the modes on the cylinder given in section \ref{expansions} change under the transformations (a)-(c). 
 Before the twist we find
\bea
\alpha^{(1)}_{A\dot A, n}&\r& {1\over 2\pi} \int_{t=ia} \p_t X_{A\dot A}(t) (z_0+t^2)^n dt
\label{lseven}\\
\alpha^{(2)}_{A\dot A, n}&\r& {1\over 2\pi} \int_{t=-ia} \p_t X_{A\dot A}(t) (z_0+t^2)^n dt
\label{leight}\\
d^{(1)+ A}_n&\r& {2^\h\over 2\pi i} \int_{t=ia} \psi^{+ A}(t) (z_0+t^2)^{n-1}t dt
\label{lnine}\\
d^{(2)+ A}_n&\r& {2^\h\over 2\pi i} \int_{t=-ia} \psi^{+ A}(t) (z_0+t^2)^{n-1}t dt
\label{lten}\\
d^{(1)- A}_n&\r&{2^\h\over 2\pi i} \int_{t=ia} \psi^{- A}(t) (z_0+t^2)^{n}  dt
\label{lel}\\
d^{(2)- A}_n&\r&{2^\h\over 2\pi i} \int_{t=-ia} \psi^{- A}(t) (z_0+t^2)^{n}  dt
\label{ltw}
\eea
After the twist we get
\bea
\alpha_{A\dot A, n}&\r& {1\over 2\pi} \int_{t=\infty} \p_t X_{A\dot A}(t) (z_0+t^2)^{{n\over 2}} dt
\label{qatwo}\\
d^{+ A}_n&\r& {2^\h\over 2\pi i}\int_{t=\infty} \psi^{+ A}(t) (z_0+t^2)^{{(n-2)\over 2} } t dt
\label{lthir}\\
d^{- A}_n&\r& {2^\h\over 2\pi i}\int_{t=\infty} \psi^{- A}(t) (z_0+t^2)^{{n\over 2} }  dt\label{lthirf}
\eea

\subsection{The final state}

Expressing the $t$ plane vacuum in terms of the cylinder modes of the original problem, we find  \cite{acm1}
\be
|\chi\rangle=e^{\sum_{ m\ge 1, n\ge 1}\gamma^B_{mn}[-\alpha_{++, -m}\alpha_{--, -n}+\alpha_{-+, -m}\alpha_{+-, -n}]}
e^{\sum_{m\ge0,n\ge 1}\gamma^F_{mn}[d^{++}_{-m}d^{--}_{-n}-d^{+-}_{-m}d^{-+}_{-n}]}|0^-_R\rangle
\label{pfive}
\ee
where
\be\label{gammaB}
\gamma^B_{2m'+1, 2n'+1}={2\over (2m'+1)(2n'+1)} {z_0^{(1+m'+n')} \Gamma[{3\over 2}+m']\Gamma[{3\over 2}+n']\over (1+m'+n')\pi \Gamma[m'+1]\Gamma[n'+1]}
\ee
\be\label{gammaF}
\gamma^{F}_{2m'+1, 2n'+1}=-{ z_0^{(1+m'+n')}\Gamma[{3\over 2}+m']\Gamma[{3\over 2}+n']\over (2n'+1)\pi(1+m'+n') \gamma[m'+1]\Gamma[n'+1]}
\ee
Finally we must apply the supercharge on $|\chi\rangle$ to obtain $|\psi\rangle$. We have
\be
G^-_{\dot A, 0}={1\over 2\pi i} \int_{w=\tau}^{w=\tau+4\pi i} G^-_{\dot A}(w) dw = -{i\over 2} \sum_{n=-\infty}^\infty d^{-A}_n \alpha_{A\dot A, -n}
\ee
and find
\bea
|\psi\rangle&=&G^-_{\dot A, 0}|\chi\rangle\nonumber\\
&=&-i\Big (\sum_{n'\ge 0} {2^\h z_0^{n'+\h}\Gamma[{3\over 2}+n'] \over \pi^\h (2n'+1)\Gamma[n'+1]}d^{-A}_{-(2n'+1)}\Big )
\Big (\sum_{m'\ge 0} {2^\h z_0^{m'+\h}\Gamma[{3\over 2}+m'] \over \pi^\h (2m'+1)\Gamma[m'+1]}\alpha_{A\dot A, -(2m'+1)}\Big )\nonumber \\
&&e^{\sum_{ p'\ge 0, q'\ge 0}\gamma^B_{2p'+1, 2q'+1}[-\alpha_{++, -(2p'+1)}\alpha_{--, -(2q'+1)}+\alpha_{-+, -(2p'+1)}\alpha_{+-, -(2q'+1)}]}\nonumber \\
&&e^{\sum_{p''\ge 0,q''\ge 0}\gamma^F_{2p''+1, 2q''+1}[d^{++}_{-(2p''+1)}d^{--}_{-(2q''+1)}-d^{+-}_{-(2p''+1)}d^{-+}_{-(2q''+1)}]}~~|0^-_R\rangle
\label{finalstate}
\eea
A similar expression is obtained for $|\bar\psi\rangle=-\bar G^-_{\dot B, 0}|\bar\chi\rangle$, and we thus find the complete state resulting from the action of the deformation operator on the spin down Ramond vacuum. 

\section{Outline of the computations}\label{method}

We now wish to consider the situation where we have an excitation in the initial state on the cylinder. These excitations are generated by operators $\alpha_{A\dot A, n}, d^{\alpha A}_n$. Here we must have $n\le -1$ for bosonic excitations, since there is no momentum and so the zero mode $\alpha_{A\dot A, 0}$ kills the vacuum. For the fermions, we have $n\le 0$ for $d^{+A}_n$ and $n<0$ for $d^{-A}_n$ since we cannot apply the zero mode $d^{-A}_0$ to the vacuum $|0_R^{--}\rangle^{(1)}\otimes |0_R^{--}\rangle^{(2)}$ that we have taken as the starting point before we apply the excitation modes.

Let us note the steps we will have to perform in general:

\b

(i) In section \ref{scharge} we decomposed the action of the supercharge into an application of $-(G^{(1)-}_{\dot A, 0}+G^{(2)-}_{\dot A, 0})$ below the twist and an application of $G^-_{\dot A,0}$ above the twist.  Let us first consider the part below the twist. We commute these supercharge modes down through any excitations   $\alpha_{A\dot A, n}, d^{\alpha A}_n$  present in the initial state. After a supercharge zero mode passes all these modes  it will reach the state $|0_R^{--}\rangle^{(1)}\otimes |0_R^{--}\rangle^{(2)}$ at the bottom of the cylinder, and by (\ref{lthree}) we will get zero. The commutation through the excitations can be computed by the relations
\be
G^{(1)-}_{\dot A,0}= -i \sum_{n=-\infty}^\infty d^{(1)-A}_{-n} \alpha^{(1)}_{A\dot A, n}
\label{kktwo}
\ee
\bea
[G^{(1)-}_{\dot A,0},\alpha^{(1)}_{B\dot B, m}]&=&
-im \epsilon_{AB}\epsilon_{\dot A\dot B}d^{(1)-A}_{m}\\
\{ G^{(1)-}_{\dot A,0}, d^{(1)-B}_{m}\}&=&0\\
\{G^{(1)-}_{\dot A,0} ,d^{(1)+B}_{m}\}&=&i\epsilon^{AB}\alpha^{(1)}_{A\dot A, m}
\eea
We have exactly similar relations  for the operators in copy $2$.

\b

(ii)  Having disposed of any supercharge present below the twist, we get the twist operator $\sigma_2^+$ acting on a set of  bosonic and fermionic modes. In the sections below we will  take the case where we have a single bosonic or fermionic mode below the twist $\sigma_2^+$, and will find that this can be expressed as a linear combination of single particle modes acting {\it after} the twist. If we have several modes below the twist, then we can perform the same process with each mode, except that there can be in addition a `Wick contraction' between a pair of bosonic or fermionic modes. This results in a c-number contribution, which we compute as well. 

\b

(iii) In this manner we obtain a state where we have $\sigma_2^+$ acting on the spin down Ramond vacuum, and a set of excitations acting after this twist. The action of $\sigma_2^+$ on the Ramond  vacuum gives the state $|\chi\rangle$ (defined in (\ref{chiq})); this state was found in \cite{acm1} and is given in (\ref{pfive}). Thus we get a set of modes acting on $|\chi\rangle$. Finally we note that we have a part of the amplitude where we must act with the supercharge mode (\ref{lfive}) applied after all other operators. This is done using the relations 
\be
G^-_{\dot A, 0}=-{i\over 2} \sum_{n=-\infty}^\infty d^{-A}_n \alpha_{A\dot A, -n}
\ee
\bea
[G^{-}_{\dot A,0},\alpha^{}_{B\dot B, m}]&=&
-\h im \epsilon_{AB}\epsilon_{\dot A\dot B}d^{-A}_{m}\\
\{ G^{-}_{\dot A,0}, d^{-B}_{m}\}&=&0\\
\{G^{-}_{\dot A,0} ,d^{+B}_{m}\}&=&i\epsilon^{AB}\alpha^{}_{A\dot A, m}
\label{kkthree}
\eea
We can use these relations to commute the supercharge down through the modes now present above $\sigma_2^+$,  until it reaches the twist insertion.  The supercharge applied to the twist, with the Ramond vacuum below the twist, results in the state $|\psi\rangle$ given in (\ref{finalstate}) \cite{acm1}.
Thus we end up with a set of modes on the doubly wound circle, acting on 
$|\chi\rangle, |\psi\rangle$. 

\b

In this manner we compute the full state resulting from the action of the deformation operator on a general initial state containing excitations.

\section{The action of $\sigma_2^+$ on a single mode}\label{single}

In this section we perform the following computation: we have a single excitation in the initial state, and the action of $\sigma_2^+$ on this state. The created state will have the same exponential as in the state $|\chi\rangle$ created by $\sigma_2^+$ from the vacuum, but in addition the initial excitation will split into a linear combination of single particle modes  in the final state. 

\subsection{The action of $\sigma_2^+(w_0)$ on a bosonic mode}

Let us find the state
\be
|\xi\rangle=\sigma_2^+(w_0)\alpha^{(1)}_{A\dot A,n}|0_R^{--}\rangle^{(1)}\otimes |0_R^{--}\rangle^{(2)}
\ee
Since the vacuum state $|0_R^{--}\rangle^{(1)}\otimes |0_R^{--}\rangle^{(2)}$ is killed by nonnegative modes of the boson, we will take 
\be
n\le -1
\label{lfourt}
\ee

We perform the spectral flows and coordinate maps given in section (\ref{steps}). This brings us to the $t$ plane with all punctures at $t=\pm ia, t=0$, smoothly closed. Before these steps, the mode $\alpha^{(1)}_{A\dot A,n}$ is given in (\ref{lone}), and after all the spectral flow and coordinate maps, it is given in the $t$ plane by (\ref{lseven})
\be
\alpha^{(1)}_{A\dot A, n}\r {1\over 2\pi} \int_{t=ia} \p_t X_{A\dot A}(t) (z_0+t^2)^n dt
\label{lsevenp}
\ee
We will now go through a sequence of steps to find the effect of this mode in the final state. The essence of these steps in the following. The initial state modes (\ref{lsevenp}) are defined by a contour around the point $t=ia$. The final state modes (\ref{qatwo}) are defined at large $t$. They involve the function $(z_0+t^2)^{{n\over 2}}$ which has branch points in the $t$ plane for odd $n$, so we cannot directly stretch the contour giving the initial state modes to get the modes in the final state. Instead, we proceed from the initial modes to the final modes through a sequence of steps that expands the initial contour into a linear combination of modes at large $t$.

\subsubsection{The operators $\hat\alpha_{A\dot A, n}$}\label{alphahat}

The contour in (\ref{lsevenp}) circles the point $t=ia$. As we saw in section (\ref{steps}), there is no singularity at $t=ia$ after all spectral flows and coordinate maps have been done. Thus if we define modes 
\be
\hat\alpha^{(1)}_{A\dot A, n}={1\over 2\pi}\int_{t=ia} dt \p_t X_{A\dot A}(t) (t-ia)^n
\label{ften}
\ee 
then we will find
\be
\hat\alpha_{A\dot A, n}|0\rangle_{ia}=0, ~~~n\ge 0
\label{ltenqq}
\ee
where $|0\rangle_{ia}$ is the NS vacuum at the point $t=ia$, and we have noted that the zero mode vanishes since there is no momentum for the boson at any stage. The modes (\ref{ften}) have the commutation relations
\be
[\hat\alpha_{A \dot A,m}, \hat\alpha_{B \dot B,n}]=-\epsilon_{A B} \epsilon_{\dot A \dot B} m\delta_{m+n, 0}
\label{pttwoqq}
\ee

Thus we would like to expand the mode (\ref{lsevenp}) in modes of type (\ref{ften}). Writing
\be
z_0+t^2=(t-ia)(t+ia)
\ee
we write the  mode (\ref{lsevenp}) as
\be
\alpha^{(1)}_{A\dot A, n}\r {1\over 2\pi} \int_{t=ia} \p_t X_{A\dot A}(t) (t-ia)^n(t+ia)^n dt
\ee
We have 
\bea
(t+ia)^n&=&\Big ( 2ia + (t-ia)\Big )^n\nn
&=&(2 i a)^n \Big ( 1+(2 i a)^{-1} (t-ia)\Big ) ^n\nn
&=&\sum_{k\ge 0} {}^n C_k (2ia)^{(n-k)} (t-ia)^k
\eea
We thus find
\be
\alpha^{(1)}_{A\dot A, n}\r \sum_{k\ge 0} {}^n C_k (2ia)^{(n-k)}{1\over 2\pi}\int_{ia} dt \p_t X_{A\dot A}(t)(t-ia)^{(n+k)}=\sum_{k\ge 0} {}^n C_k (2ia)^{(n-k)} \hat \alpha^{(1)}_{n+k}
\label{ktwo}
\ee
From  (\ref{lten})   we find that we will get a nonzero contribution only from modes with
\be
n+k\le -1~~\Rightarrow~~k\le -n-1
\ee
(Recall from (\ref{lfourt}) that $n$ is negative.) Thus we have
\be
\alpha^{(1)}_{A\dot A, n}\r \sum_{k=0}^{-n-1} {}^n C_k (2ia)^{(n-k)} \hat \alpha^{(1)}_{n+k}
\label{lsixt}
\ee
Note that this is a finite sum; this is important to have convergence for sums that we will encounter below.

\subsubsection{Converting to modes $\tilde\alpha_{A\dot A, q}$}

The contour in the operators $\hat\alpha_{A\dot A, n}$ circles $t=ia$. But as we have seen in section (\ref{steps}), there are no  singularities at other points in the $t$ plane, so we can stretch this contour into a contour at large $t$, which we write as $\int_{t=\infty}$. We get
\be
\hat\alpha^{(1)}_{A\dot A, m}\r {1\over 2\pi}\int_{t=\infty} dt \p_t X_{A\dot A}(t) (t-ia)^m
\ee
\be
(t-ia)^m=t^m\Big (1-(ia)t^{-1}\Big )^m=\sum_{k'\ge 0} {}^mC_{k'}(-ia)^{k'} t^{m-k'}
\ee
Let us now define modes natural to large $t$ in the $t$ plane:
\be
\t\alpha_{A\dot A, m}\equiv  {1\over 2\pi} \int_{t=\infty} \p_t X_{A\dot A}(t) t^m dt
\label{qathreeq}
\ee
We have
\be
[\t\alpha_{A \dot A}, \t\alpha_{B \dot B}]=-\epsilon_{A B} \epsilon_{\dot A \dot B} m\delta_{m+n, 0};~ ~~~\t\alpha_{A \dot A,m}|0\rangle_t=0, ~~~m\ge 0
\label{ptthree}
\ee
We find
\be
\hat\alpha_{A\dot A, m} = \sum_{k'\ge 0} {}^mC_{k'}(-ia)^{k'}\t \alpha_{A\dot A, m-k'}
\ee

Let us now substitute this expansion for $\hat\alpha_{A\dot A, m}$ into (\ref{lsixt}). We get
\be
\alpha^{(1)}_{A\dot A, n}\r \sum_{k= 0}^{-n-1} {}^n C_k (2ia)^{(n-k)}\sum_{k'\ge 0} {}^{(n+k)}C_{k'}(-ia)^{k'}\t \alpha_{A\dot A, n+k-k'}
\ee
Let us look at the coefficient of $\t\alpha_{A\dot A, q}$. Thus 
\be
q=n+k-k' ~~\Rightarrow ~~k'=n+k-q
\ee
Since we have $k'\ge 0$, we find
\be
n+k-q\ge 0 ~~\Rightarrow~~k\ge q-n
\ee
Thus the $k$ sum runs over the following range
\bea
q-n\le 0:&~~&k=[0, -n-1]\nn
q-n\ge0:&~~&k=[q-n, -n-1]
\eea
Recall that $n\le -1$. The second of these equations tells us that there is no summation range at all for $q\ge 0$. Thus we only generate negative index modes $\t\alpha_{A\dot A, q}$ from our expansion:
\be
q\le -1
\label{qequation}
\ee
We now have
\bea
\alpha^{(1)}_{A\dot A, n}&\r& \sum_{q\le -1} \Big (\sum_{k= 0}^{-n-1} {}^n C_k (2ia)^{(n-k)} ~{}^{(n+k)}C_{n+k-q}(-ia)^{n+k-q}\Big )~\t \alpha_{A\dot A, q}, ~~~q\le n\nn
\alpha^{(1)}_{A\dot A, n}&\r& \sum_{q\le -1} \Big (\sum_{k= q-n}^{-n-1} {}^n C_k (2ia)^{(n-k)} ~{}^{(n+k)}C_{n+k-q}(-ia)^{n+k-q}\Big )~\t \alpha_{A\dot A, q}, ~~~q\ge n\nn
\eea
The sum has the same algebraic expression in both the ranges for $q$, and we find
\be
\alpha^{(1)}_{A\dot A, n}\r \sum_{q\le -1} \Big (    { i^{-q}(-1)^n  a^{2n-q}  \Gamma[-{q\over 2}] \over 2 \Gamma[-n]\Gamma[n+1-{q\over 2}]}
\Big )~\t \alpha_{A\dot A, q}
\label{ltwthree}
\ee

\subsubsection{Converting to modes $\alpha_{A\dot A, p}$}

We must finally convert to the modes on the cylinder at $\tau\r\infty$. These modes are given in (\ref{qatwo})
\be
\alpha_{A\dot A, p}\r {1\over 2\pi} \int_{t=\infty} dt \p_t X_{A\dot A}(t) (z_0+t^2)^{{p\over 2}} 
\label{ltwone}
\ee
Write
\be
t'=(t^2+z_0)^\h~~\Rightarrow ~~ t=(t'^2-z_0)^\h
\ee
with the sign of the square root chosen to give $t'\sim t$ near infinity. Then we have
\be
t^q=(t'^2-z_0)^{q\over 2}=t'^q\Big (1-z_0 t'^{-2}\Big)^{q\over 2}=\sum_{k\ge 0} {}^{q\over 2}C_k (-z_0)^k t'^{q-2k}=\sum_{k\ge 0} {}^{q\over 2}C_k (-z_0)^k (z_0+t^2)^{q-2k\over 2}
\ee
Substituting this expansion in (\ref{qathreeq}) we find that
\be
\t\alpha_{A\dot A, q}={1\over 2\pi}\int dt \p_{t}X_{A\dot A}(t)\Big (\sum_{k\ge 0} {}^{q\over 2}C_k (-z_0)^k (z_0+t^2)^{q-2k\over 2} \Big )
\ee
Using (\ref{ltwone}) we find 
\be
\t\alpha_{A\dot A, q}=\sum_{k\ge 0} {}^{q\over 2}C_k (-z_0)^k\alpha_{A\dot A, q-2k}
\label{ltwtwo}
\ee

We substitute the expansion (\ref{ltwtwo}) into (\ref{ltwthree}), getting
\be
\alpha^{(1)}_{A\dot A, n}\r \sum_{q\le -1} \Big (    { i^{-q}(-1)^n  a^{2n-q}  \Gamma[-{q\over 2}] \over 2 \Gamma[-n]\Gamma[n+1-{q\over 2}]}
\Big )~\sum_{k\ge 0} {}^{q\over 2}C_{k} (-z_0)^{k}\alpha_{A\dot A, q-2k}
\label{ltthree}
\ee
Let us look at the coefficient of $\alpha_{A\dot A, p}$ in this sum. This gives
\be
p=q-2k
\label{ltwfive}
\ee
Note that since $q\le -1$ and $k\ge 0$, we have 
\be
p\le -1
\label{ltwfour}
\ee
From (\ref{ltwfive}) we see that if $p$ is even only even values of $q$ contribute to this sum, and if $p$ is odd then only odd values of $q$ contribute. 
For the even $p$ case we write $p=2p', q=2q'$.
From (\ref{ltwfive}) we set $k=q'-p'$. 
Since $k\ge 0$, the range of the $q'$ sum becomes $p'\le q'\le -1$. We get the sum
\be
\sum_{q'=p' }^{-1} { (-1)^{q'+n}  a^{2n-2q'}  \Gamma[-{q'}] \over 2 \Gamma[-n]\Gamma[n+1-q']}
~ {}^{q'}C_{q'-p'} (-z_0)^{q'-p'}=\h\delta_{n, p'}
\ee
For odd $p$ we write $p=2p'+1, q=2q'+1$. From (\ref{ltwfive}) we  again get $k=q'-p'$.
Since $k\ge 0$, the range of $q'$ is $p'\le q'\le -1$. We get the sum
\bea
\sum_{q'=p' }^{-1} &&{ i^{-1}(-1)^{q'+n}  a^{2n-2q'-1}  \Gamma[-{q'}-\h] \over 2 \Gamma[-n]\Gamma[n+\h-q']}
~ {}^{q'+\h}C_{q'-p'} (-z_0)^{q'-p'}\nn
&&~~~~~~~~~=~~{i\over \pi} {\Gamma[\h-n]\over \Gamma[-n]}{\Gamma[-\h-p']\over \Gamma[-p']}{a^{2(n-p')-1}\over (2n-2p'-1)}
\eea

\subsubsection{Summary}

Now that we have converted the initial mode $\alpha_{A\dot A, n}^{(1)}$ to modes at large $t$, we note that we are left with the NS vacuum $|0\rangle_t$ of the $t$ plane inside these modes. This vacuum just gave us the state $|\chi\rangle$ on the cylinder \cite{acm1}
\be
\sigma_2^+(w_0) |0_R^{--}\rangle^{(1)}\otimes |0_R^{--}\rangle^{(2)}=|\chi\rangle
\ee

Putting the above results together, we see that for $n<0$
\bea
\hskip -.3 in  \sigma_2^+(w_0)\alpha^{(1)}_{A\dot A, n} |0_R^{--}\rangle^{(1)}&&\hskip -.3 in\otimes |0_R^{--}\rangle^{(2)}\nn
=\Bigg (\h \alpha_{A\dot A, 2n}+\sum_{p'\le -1}&&\hskip -.2 in\Big ({i\over \pi} {\Gamma[\h-n]\over \Gamma[-n]}{\Gamma[-\h-p']\over \Gamma[-p']}{a^{2(n-p')-1}\over (2n-2p'-1)}
\Big ) ~\alpha_{A\dot A, 2p'+1}\Bigg )|\chi\rangle\nn
\hskip -.1 in\equiv ~\sum_p f^B[n,p]\alpha_{A\dot A, p}&&\hskip -.2 in|\chi\rangle
\label{kkone}
\eea
where we have defined the coefficients $f^B[n,p]$ for later convenience. 
For $n\ge 0$ we will just have
\be
 \sigma_2^+(w_0)\alpha^{(1)}_{A\dot A, n}  |0_R^{--}\rangle^{(1)}\otimes |0_R^{--}\rangle^{(2)}=0
\ee
since positive modes annihilate the Ramond vacuum state.

We see that the even modes in the final state get a simple contribution; this is related to the fact that the twist operator $\sigma_2$ does not affect such modes when it cuts and joins together the two copies of the $c=6$ CFT. The odd modes of all levels are excited.

\subsection{The action of $\sigma_2^+(w_0)$ on a fermionic mode}

Let us repeat the above computation for a fermionic mode in the initial state. 
There is a slight difference between the cases of $d^{-A}_n$ and $d^{+A}_n$ since the starting vacuum state $|0_R^{--}\rangle^{(1)}\otimes |0_R^{--}\rangle^{(2)}$  breaks the charge symmetry, and the spectral flows we do also break this symmetry.

\subsubsection{The mode $d^{-A}_n$ in the initial state}

We start with eq. (\ref{lel}), which we write as
\be
d^{(1)-A}_n\r {2^\h\over 2\pi i} \int_{ia} dt~ \psi^{-A}_t(t)(t-ia)^n(t+ia)^n
\ee
We perform the spectral flows and coordinate maps in section (\ref{steps}) to reach the $t$ plane with all punctures smoothly closed. 
Define natural modes around the point $t=ia$ in the $t$ plane
\be
\hat d^{-A}_r={1\over 2\pi i}\int_{ia} dt~ \psi^{-A}_t(t) (t-ia)^{r-\h}
\ee
where $r$ is a half-integer. Writing $t+ia=2ia+(t-ia)$, we expand in powers of $(t-ia)$. Noting that  operators $\hat d^{-A}_r$ with $r>0$ kill the NS vacuum at $t=ia$, we find
\be
d^{(1)-A}_n~\r~ 2^\h \sum_{k=0}^{-n-1}~{}^nC_k~ (2ia)^{n-k} \hat d^{-A}_{n+k+\h}
\ee

The RHS in the above equation is a finite sum of operators, each given by a contour integral around $t=ia$. Since there are no singularities anywhere on the $t$ plane, we can expand each contour to one at large $t$. We define operators natural for expansion around infinity in the $t$ plane
\be
\t d^{-A}_r={1\over 2\pi i}\int_{t=\infty} dt~ \psi^{-A}_t(t)~ t^{r-\h}
\ee
where $r$ is a half-integer. The commutation relations are
\be
\{\t d^{\alpha A}_r, \t d^{\beta B}_s\}=-\epsilon^{\alpha\beta}\epsilon^{AB}\delta_{r+s, 0}
\label{pttwop}
\ee
We find
\be
\hat d^{-A}_r=\sum_{k'\ge 0} ~{}^{r-\h}C_{k'} ~(-ia)^{k'}~\t d^{-A}_{r-k'}
\ee

Finally we can expand the operators $\t d^{-A}_r$ in terms of the final state modes (\ref{lthirf}), finding
\be
\t d^{-A}_r=2^{-\h} \sum_{k\ge 0} ~{}^{r-\h\over 2}C_k ~ (-a^2)^k ~d^{-A}_{r-2k-\h}
\ee

We then find
\bea
\hskip -.3 in  \sigma_2^+(w_0) d^{(1)-A}_n |0_R^{--}\rangle^{(1)}&&\hskip -.3 in\otimes |0_R^{--}\rangle^{(2)}\nn
=\Bigg (\h d^{-A}_{2n}+\sum_{p'\le -1}&&\hskip -.2 in\Big ({i\over \pi} {\Gamma[\h-n]\over \Gamma[-n]}{\Gamma[-\h-p']\over \Gamma[-p']}{a^{2(n-p')-1}\over (2n-2p'-1)}
\Big ) ~d^{-A}_{2p'+1}
\Bigg )|\chi\rangle\nn
\hskip -.1 in\equiv ~\sum_p f^{F-}[n,p]&&\hskip -.2 ind^{-A}_p|\chi\rangle
\label{jtwo}
\eea

\subsubsection{The mode $d^{+A}_n$ in the initial state}

We start with (\ref{lnine}), which we write as
\be
d^{(1)+A}_n\r {2^\h\over 2\pi i} \int_{ia} dt~ \psi^{+A}_t(t)(t-ia)^{n-1}(t+ia)^{n-1}~t
\label{jone}
\ee
We perform the steps in section (\ref{steps}) as before.  The natural modes around $t=ia$ are
\be
\hat d^{+A}_r={1\over 2\pi i}\int_{ia} dt~ \psi^{+A}_t(t) (t-ia)^{r-\h}
\ee
This time we must expand in (\ref{jone}) the factor $t+ia=2ia+(t-ia)$ as well as the factor $t=ia + (t-ia)$. Thus generates two terms
\be
d^{(1)+A}_n~\r~{1\over 2^\h} \sum_{k= 0}^{-n} {}^{n-1}C_k (2ia)^{n-k}\hat d^{+A}_{n+k-\h}+2^\h \sum_{k= 0}^{-n-1} {}^{n-1}C_k (2ia)^{n-k-1}\hat d^{+A}_{n+k+\h}
\ee
Define natural modes at large $t$
\be
\t d^{+A}_r={1\over 2\pi i}\int_{t=\infty} dt~ \psi^{+A}_t(t)~ t^{r-\h}
\ee
We find
\be
\hat d_r^{+A}~=~\sum_{k\ge 0}{}^{r-\h}C_k (-ia)^k \t d^{+A}_{r-k}
\ee
Finally we can expand the modes $\t d^{+A}$ in terms of the final state modes (\ref{lthir}), finding
\be
\t d_r^{+A}=2^{-\h}\sum_{k\ge 0} ~{}^{r-{3\over 2}\over 2}C_k ~(-a^2)^k~d^{+A}_{r-2k+\h}
\ee
Putting all these expansions together, we find
\bea
\hskip -.3 in  \sigma_2^+(w_0)d^{(1)+A}_n |0_R^{--}\rangle^{(1)}&&\hskip -.3 in\otimes |0_R^{--}\rangle^{(2)}\nn
=\Bigg (\h d^{+A}_{2n}+\sum_{p'\le -1}&&\hskip -.2 in\Big ({i\over \pi} {\Gamma[\h-n]\over \Gamma[1-n]}{\Gamma[\h-p']\over \Gamma[-p']}{a^{2(n-p')-1}\over (2n-2p'-1)}
\Big ) ~d^{+A}_{2p'+1}
\Bigg )|\chi\rangle\nn
\hskip -.1 in\equiv ~\sum_p f^{F+}[n,p]&&\hskip -.2 ind^{+A}_p|\chi\rangle
\eea

\subsection{Summary}

The computations of this section are are very basic to understanding the effect of the deformation operator: taken by itself, any single particle mode below the twist insertion $\sigma_2^+(w_0)$ spreads into a linear combination of single particle modes after the twist, and we have found the coefficients of this linear combination for both bosonic and fermionic excitations. In addition the twist creates the same exponential that arises in the action of the twist of the vacuum (eq. (\ref{pfive}), so the action of the twist on a single particle initial state gives rise to states with $1,3,5,\dots$ excitations.

\section{Two modes in the initial state}\label{twomodes}

Now let us consider the situation where we have two excitations in the initial twist. Upon the action of the twist $\sigma_2^+$ there will be two kinds of terms. One, where the modes move separately to the final state; this contribution can thus be computed by using the expressions of the last section. The other contribution results from an interaction between the two modes. Since we are dealing with a theory of free bosons and free fermions, 
the only possible interactions between modes is a `Wick contraction', which produces a c-number term. 

\subsection{Two bosonic modes}

Let us consider the state
\be
 \sigma_2^+(w_0)\alpha^{(1)}_{A\dot A, n_1}\alpha^{(1)}_{B\dot B, n_2}  |0_R^{--}\rangle^{(1)}\otimes |0_R^{--}\rangle^{(2)}
\label{kthree}
\ee
with $n_1<0, n_2<0$. We wish to move the modes $\alpha^{(1)}_{A\dot A, n_i}$ to operator modes acting after the $\sigma_2^+$ operator. We will get the terms corresponding to each of these modes moving across separately, but there will also be a term resulting from the interaction between the two modes.

We follow the sequence of spectral flows and dualities given in section (\ref{steps}). We reach the $t$ plane with all punctures closed and the operator modes (cf. eq. (\ref{lseven}))
\be
\alpha^{(1)}_{A\dot A, n_1}\alpha^{(1)}_{B\dot B, n_2}\r\Big ({1\over 2\pi} \int_{t=ia} \p_t X_{A\dot A}(t_1) (z_0+t_1^2)^{n_1} dt_1\Big )\Big ({1\over 2\pi} \int_{t=ia} \p_t X_{B\dot B}(t_2) (z_0+t_2^2)^{n_2} dt_2\Big )
\ee
with the $t_1$ contour outside the $t_2$ contour.

There is no singularity inside the $t_2$ contour, so we can expand the $t_2$ contour as in section (\ref{alphahat}) to get (cf. eq. (\ref{lsixt}))
\be
\alpha^{(1)}_{B\dot B, n_2}\r \sum_{k_2=0}^{-n_2-1} {}^{n_2} C_{k_2} (2ia)^{(n_2-k_2)} \hat \alpha^{(1)}_{B\dot B, n_2+k_2}
\label{kone}
\ee
For the $t_1$ contour we can get a contribution from both positive and negative $\hat\alpha$ modes, since the $t_2$ contour gives an operator inside the $t_1$ contour. Thus we write the general expansion (\ref{ktwo}) for this contour 
\be
\alpha^{(1)}_{A\dot A, n_1}\r \sum_{k_1=0}^{\infty} {}^{n_1} C_{k_1} (2ia)^{(n_1-k_1)}~ \hat \alpha^{(1)}_{A\dot A, n_1+k_1}
\label{konef}
\ee
and consider separately two cases:

\b

(a) The range of $k_1$ where $n_1+k_1\le -1$. This gives negative index modes just like (\ref{kone}). These modes commute with the modes in (\ref{kone}), so we have no interaction between the two operators, and we get (cf. (\ref{lsixt}))
\be
\alpha^{(1)}_{A\dot A, n_1}\r \sum_{k_1=0}^{-n_1-1} {}^{n_1} C_{k_1} (2ia)^{(n_1-k_1)}~ \hat \alpha^{(1)}_{A\dot A, n_1+k_1}
\label{koneq}
\ee

\b

(b) The range where $n_1+k_1\ge 0$. Now these modes can annihilate the negative modes created by the $t_2$ contour. This results in a c-number contribution
\be
C^B_{A\dot A B\dot B}[n_1,n_2]=\sum_{k_1=-n_1}^{\infty} \sum_{k_2=0}^{-n_2-1} ~{}^{n_1} C_{k_1} (2ia)^{(n_1-k_1)} ~{}^{n_2} C_{k_2} (2ia)^{(n_2-k_2)}~[\hat \alpha^{(1)}_{A\dot A, n_1+k_1}, \hat \alpha^{(1)}_{B\dot B, n_2+k_2}]
\ee
Using the commutation relation (\ref{pttwoqq}) we get
\bea
C^B_{A\dot A B\dot B}[n_1,n_2]&=&(-\epsilon_{A B} \epsilon_{\dot A \dot B})\Big (\sum_{k_2=0}^{-n_2-1} (-(n_2+k_2))~{}^{n_1} C_{-n_1-n_2-k_2}  ~{}^{n_2} C_{k_2}~(2ia)^{2(n_1+n_2)} \Big)\nn
&=&\epsilon_{A B} \epsilon_{\dot A \dot B}\Big ({ a^{2(n_1+n_2)}\Gamma[-n_1+\h]\Gamma[-n_2+\h]\over 2\pi (n_1+n_2) \Gamma[-n_1]\Gamma[-n_2]} \Big )
\label{wickb}
\eea
Note that this is symmetric in $n_1$ and  $n_2$, as it should be since the modes in (\ref{kthree}) commute and so can be put in either order. 

Apart from this c-number term, we still have the contribution where the two contours in the initial state produce the modes (\ref{kone}),(\ref{koneq}). We can proceed to expand the modes $\hat\alpha$ in these sums into modes of type $\t\alpha$. We note from (\ref{qequation}) that each set of modes generates only negative index modes $\t\alpha_q$. Thus we cannot get any additional c-number contributions from commutators between the $\t\alpha$ modes arising from our two operators.

Next we convert the modes $\t\alpha$ to modes of type $\alpha$. From (\ref{ltwfour}) we see that we again generate only negative index modes $\alpha_p$ from each of the two sets of modes. Thus  there cannot be any additional c-number contribution from commutation between the modes $\alpha$ arising from our two operators. In short, the only c-number contribution we get from `Wick contraction' is (\ref{wickb}), and the remaining part of the state is given by independently moving the two initial state modes past $\sigma_2^+$ in the manner given in (\ref{kkone}).

Putting this `Wick contraction' term together with the contribution of the uncontracted terms  we get
\bea
\sigma_2^+(w_0) \alpha^{(1)}_{A\dot A, n_1}\alpha^{(1)}_{B\dot B, n_2} | 0_R^{--}\rangle^{(1)}&&\hskip -.3 in\otimes |0_R^{--}\rangle^{(2)}\nn
=\Big [\Big (\sum_{p_1} f^B[n_1,p_1]\alpha_{A\dot A, p_1}\Big )&&\hskip -.3 in\Big (\sum_{p_2} f^B[n_2,p_2]\alpha_{B\dot B, p_2}\Big )+C^B_{A\dot A B\dot B}[n_1,n_2]\Big ] ~|\chi\rangle
\eea

\subsection{Two fermionic modes}

Let us repeat this computation for two fermions in the initial state. Since the spectral flow treats positive and negative charges differently, we work with the pair $d^{++}, d^{--}$ and later write the result for general charges.

Consider 
\be
\sigma_2^+(w_0) d^{(1),++}_{ n_1}d^{(1),--}_{ n_2} |0_R^{--}\rangle^{(1)}\otimes |0_R^{--}\rangle^{(2)}
\label{kthreeq}
\ee
with $n_1\le0, n_2<0$. We wish to move the modes to those acting after the $\sigma_2^+$ operator. We will get the terms corresponding to each of these modes moving across separately, but there will also be a term resulting from the interaction between the two modes.

Following the sequence of spectral flows and dualities we reach the $t$ plane with all punctures closed and the operator modes (cf. eq.(\ref{lnine}),(\ref{lel}))
\be
d^{(1),++}_{ n_1}d^{(1),--}_{ n_2} \r\Big ({1\over 2\pi i} 2^\h\int_{t=ia} \psi_t^{++}(t_1) (z_0+t_1^2)^{n_1-1}t_1 dt_1\Big )\Big ({1\over 2\pi i} 2^\h\int_{t=ia} \psi_t^{--}(t_2) (z_0+t_2^2)^{n_2} dt_2\Big )
\ee
with the $t_1$ contour outside the $t_2$ contour.

There is no singularity inside the $t_2$ contour so we get
\be
d^{(1),--}_{ n_2}\r 2^\h\sum_{k_2=0}^{-n_2-1} ~{}^{n_2}C_{k_2}~ (2ia)^{n_2-k_2}~{\hat d}^{--}_{n_2+k_2+\h}
\ee
For $d^{(1),++}_{ n_1}$ we write the full sum over modes
\be
d^{(1)+A}_{n_1}~\r~{1\over 2^\h} \sum_{k_1= 0}^{\infty} {}^{n_1-1}C_{k_1} (2ia)^{n_1-k_1}\hat d^{+A}_{n_1+k_1-\h}+2^\h \sum_{k_1= 0}^{\infty} {}^{n_1-1}C_{k_1} (2ia)^{n_1-k_1-1}\hat d^{+A}_{n_1+k_1+\h}
\ee
 The anticommutator arising  from the first term  gives
\be
-(2ia)^{2(n_1+n_2)}\sum_{k_2=0}^{-n_2-1} {}^{n_1-1}C_{-n_1-n_2-k_2}~{}^{n_2}C_{k_2}
\ee
while the anticommutator from the second term gives
\be
-2(2ia)^{2(n_1+n_2)}\sum_{k_2=0}^{-n_2-1} {}^{n_1-1}C_{-n_1-n_2-k_2-1}~{}^{n_2}C_{k_2}
\ee

The sum of these two contributions can be simplified to give (we now include the result for the pair $d^{+-}, d^{-+}$)
\bea
C^{F,\alpha A  \beta B}[n_1,n_2]&=&-\epsilon^{\alpha\beta}\epsilon^{A B} \Big ({ a^{2(n_1+n_2)}\Gamma[-n_1+\h]\Gamma[-n_2+\h]\over 2\pi n_1(n_1+n_2) \Gamma[-n_1]\Gamma[-n_2]} \Big )
\eea
Note that this is not symmetric in $n_1,n_2$ since the choice of Ramond vacuum $|0_R^{--}\rangle^{(1)}\otimes |0_R^{--}\rangle^{(2)}$ breaks the symmetry between $+$ and $-$ charge fermions.

Putting this `Wick contraction' term together with the contribution of the uncontracted terms we get
\bea
\sigma_2^+(w_0) d^{(1),++}_{ n_1}d^{(1),--}_{ n_2}  | 0_R^{--}\rangle^{(1)}&&\hskip -.3 in\otimes |0_R^{--}\rangle^{(2)}\nn
=\Big [\Big (\sum_{p_1} f^{F+}[n_1,p_1]d^{++}_{p_1}\Big )&&\hskip -.3 in\Big (\sum_{p_2} f^{F-}[n_2,p_2]d^{--}_{p_2}\Big )+C^{F,++--}[n_1,n_2]\Big ]~ |\chi\rangle
\eea

\subsection{Summary}

We have computed the c-number `Wick contraction' term that results from the interaction between two initial state modes. Note that after 
all the spectral flows we perform, we are dealing with a theory of free bosons and fermions. Thus even if we had several modes in the initial state, we can break up the effect of the twist $\sigma_2^+$ into pairwise `Wick contractions' (with value given by $C^B, C^F$ computed above), and moving any noncontracted modes past the twist $\sigma_2^+$ using the expressions in section (\ref{single}). So the computations of this section (\ref{single}) and the present section allow us to find the effect of $\sigma_2^+$ on any initial state.

\section{Complete action of the deformation operator on an initial bosonic mode}

In the last two sections we have focused on the effect of the twist $\sigma_2^+$. Let us now compute an example where we combine this with the action of the supercharge described in section \ref{method}. 

We start with the state containing one bosonic excitation, and find the state created by the action of the deformation operator. Thus we wish to find the state
\be
|\psi\rangle_f =\hat O_{\dot A}~\alpha_{C\dot C, n}^{(1)}~|0_R^{-}\rangle^{(1)}\otimes |0_R^{-}\rangle^{(2)}
\ee
with $n\le -1$.

We follow the steps outlined in section \ref{method}. 
We have 
\be
|\psi\rangle_f=\Big (-\sigma_2^+(w_0)~G^{(1)-}_{\dot A, 0}~\alpha_{C\dot C, n}^{(1)}
~+~G^-_{\dot A,0}~\sigma_2^+(w_0)~\alpha_{C\dot C, n}^{(1)}\Big )
|0_R^{-}\rangle^{(1)}\otimes |0_R^{-}\rangle^{(2)}
\ee
We have
\bea
\sigma_2^+(w_0)~G^{(1)-}_{\dot A, 0}~\alpha_{C\dot C, n}^{(1)}&&\hskip -.3 in|0_R^{-}\rangle^{(1)}\otimes |0_R^{-}\rangle^{(2)}\nn
&=&\sigma_2^+(w_0)~[G^{(1)-}_{\dot A, 0},\alpha_{C\dot C, n}^{(1)}]|0_R^{-}\rangle^{(1)}\otimes |0_R^{-}\rangle^{(2)}\nn
&=&
(-\epsilon_{AC}\epsilon_{\dot A\dot C})~i n~\sigma_2^+(w_0)~d^{(1)-A} _n|0_R^{-}\rangle^{(1)}\otimes |0_R^{-}\rangle^{(2)}
\eea
We can now write down $\sigma_2^+(w_0)d^{(1)-A} _n|0_R^{-}\rangle^{(1)}\otimes |0_R^{-}\rangle^{(2)}$ from (\ref{jtwo}).  For the other term, we first compute $\sigma_2^+(w_0)\alpha_{C\dot C, m}^{(1)}
|0_R^{-}\rangle^{(1)}\otimes |0_R^{-}\rangle^{(2)}$ from (\ref{kkone}), and apply the operator $G^-_{\dot A, 0}$ using 
\bea
G^-_{\dot A, 0}\alpha_{C\dot C, p}\sigma_2^+&&\hskip -.3 in|0_R^{-}\rangle^{(1)}\otimes |0_R^{-}\rangle^{(2)}\nn
&=&[G^-_{\dot A, 0},\alpha_{C\dot C, p}]\sigma_2^+|0_R^{-}\rangle^{(1)}\otimes |0_R^{-}\rangle^{(2)}+\alpha_{C\dot C, p}G^-_{\dot A,0}\sigma_2^+|0_R^{-}\rangle^{(1)}\otimes |0_R^{-}\rangle^{(2)}\nn
&=&(-\epsilon_{AC}\epsilon_{\dot A\dot C})~{ip\over 2}~d^{-A}_p~|\chi\rangle+
\alpha_{C\dot C, p}|\psi\rangle
\eea
where $|\psi\rangle$ is given in (\ref{finalstate}). 

Putting together these two contributions, we get
\bea
\hat O_{\dot A}~\alpha_{C\dot C, n}^{(1)}&&\hskip -.3 in|0_R^{-}\rangle^{(1)}\otimes |0_R^{-}\rangle^{(2)}\nn
&=&(-\epsilon_{AC}\epsilon_{\dot A\dot C})\sum_{p'\le -1}\Big ({a^{2n+1}\Gamma[\h-n]\over \pi\Gamma[-n]}\Big )\Big ({a^{-(2p'+2)}\Gamma[-p'-\h]\over 2\Gamma[-p']}\Big )d^{-A}_{2p'+1}|\chi\rangle\nn
&+&\h \alpha_{C\dot C,2n}|\psi\rangle
\eea
where the states $|\chi\rangle, |\psi\rangle$ are given in (\ref{pfive}),(\ref{finalstate}).

\section{Discussion}

In this paper we have considered several cases of the action of the deformation operator. 

We first noted that we can break up the action of the deformation operator into that of the twist $\sigma_2^+$ and that of the supercharge. The effect of the supercharge can be computed on any set of modes  using the relations (\ref{kktwo})-(\ref{kkthree}). 

The effect of the twist on a single bosonic or single fermionic mode was computed in section \ref{single}. This computation required us to relate modes defined by contours around one point $t=ia$ in the covering space, to modes defined on contours around $t=\infty$. But the function on the latter contour had branch points in the $t$ plane, and so we followed a sequence of steps that related the mode on the initial contour to a linear combination of modes on the  outer contour. 

If we have more than one mode in the initial state then we can have `Wick contractions' between these modes, and the corresponding c-number contributions were computed in section \ref{twomodes}. Taken together, these computations give a set of methods by which we can in principle compute the action of the deformation operator on any initial state. 

The overall structure of the final state is the following. Since the exponential term in (\ref{pfive}) is present in all final states obtained by the action of the deformation operator, we can get an arbitrary number of excitation pairs in the final state. Besides these pairs, we can have a mode resulting from the a mode in the initial state, and also one bosonic and one fermionic mode from the action of the supercharge $G^-$. 
Any of these modes can contract with another mode to give a c-number term; it can be seen that adding over all such possible contractions gives the final state.

\section*{Acknowledgements}

We thank Justin David for several helpful discussions. We also thank Sumit Das,  Antal Jevicki, Yuri Kovchegov, Oleg Lunin and  Emil Martinec for many helpful comments. The work of SGA and SDM is
supported in part by DOE grant DE-FG02-91ER-40690. The work of BDC was supported by the Foundation for Fundamental Research on Matter.

\appendix
\renewcommand\theequation{\thesection.\arabic{equation}}
\setcounter{equation}{0}

\section{Notation and the CFT algebra} \label{ap:CFT-notation}

We have 4 real left moving fermions $\psi^1, \psi^2, \psi^3, \psi^4$ which we group into doublets $\psi^{\alpha A}$
\be
\pmatrix{\psi^{++}\cr \psi^{-+}\cr}=\sqi\pmatrix{\psi_1+i\psi_2\cr \psi_3+i\psi_4\cr}
\ee
\be
\pmatrix{\psi^{+-}\cr \psi^{--}\cr}=\sqi\pmatrix{\psi_3-i\psi_4\cr -(\psi_1-i\psi_2)\cr}
\ee
Here $\alpha=(+,-)$ is an index of the subgroup $SU(2)_L$ of rotations on $S^3$ and $A=(+,-)$ is an index of the subgroup $SU(2)_1$ from rotations in $T^4$. The 2-point functions are
\be
<\psi^{\alpha A}(z)\psi^{\beta B}(w)>=-\epsilon^{\alpha\beta}\epsilon^{AB}{1\over z-w}
\ee
where we have defined the $\epsilon$ symbol as
\be
\epsilon_{+-}=1, ~~~\epsilon^{+-}=-1
\ee
There are 4 real left moving bosons $X_1, X_2, X_3, X_4$ which can be grouped into a matrix 
\be
X_{A\dot A}= \sqi X_i \sigma_i=\sqi\pmatrix { X_3+iX_4& X_1-iX_2\cr X_1+iX_2&-X_3+iX_4\cr}
\ee
where $\sigma_i=\sigma_a, iI$. The 2-point functions are
\be
<\p X_{A\dot A}(z) \p X_{B\dot B}(w)>={1\over (z-w)^2}\epsilon_{AB}\epsilon_{\dot A\dot B}
\ee

The chiral algebra is generated by the operators
\be
J^a=-{1\over 4}(\psi^\dagger)_{\alpha A} (\sigma^{Ta})^\alpha{}_\beta \psi^{\beta A}
\ee
\be
G^\alpha_{\dot A}= \psi^{\alpha A} \p X_{A\dot A}, ~~~(G^\dagger)_{\alpha}^{\dot A}=(\psi^\dagger)_{\alpha A} \p (X^\dagger)^{A\dot A}
\ee
\be
T=-{1\over 2} (\p X^\dagger)^{A\dot A}\p X_{A\dot A}-{1\over 2} (\psi^\dagger)_{\alpha A} \p \psi^{\alpha A}
\ee
\be
(G^\dagger)_{\alpha}^{\dot A}=-\epsilon_{\alpha\beta} \epsilon^{\dot A\dot B}G^\beta_{\dot B}, ~~~~G^{\alpha}_{\dot A}=-\epsilon^{\alpha\beta} \epsilon_{\dot A\dot B}(G^\dagger)_\beta^{\dot B}
\ee
These operators generate the algebra
\be
J^a(z) J^b(w)\sim \delta^{ab} {\h\over (z-w)^2}+i\epsilon^{abc} {J^c\over z-w}
\ee
\be
J^a(z) G^\alpha_{\dot A} (z')\sim {1\over (z-z')}\h (\sigma^{aT})^\alpha{}_\beta G^\beta_{\dot A}
\ee
\be
G^\alpha_{\dot A}(z) (G^\dagger)^{\dot B}_\beta(z')\sim -{2\over (z-z')^3}\delta^\alpha_\beta \delta^{\dot B}_{\dot A}- \delta^{\dot B}_{\dot A}  (\sigma^{Ta})^\alpha{}_\beta [{2J^a\over (z-z')^2}+{\p J^a\over (z-z')}]
-{1\over (z-w)}\delta^\alpha_\beta \delta^{\dot B}_{\dot A}T
\ee
\be
T(z)T(z')\sim {3\over (z-z')^4}+{2T\over (z-z')^2}+{\p T\over (z-z')}
\ee
\be
T(z) J^a(z')\sim {J^a\over (z-z')^2}+{\p J^a\over (z-z')}
\ee
\be
T(z) G^\alpha_{\dot A}\sim {{3\over 2}G^\alpha_{\dot A}\over (z-z')^2}  + {\p G^\alpha_{\dot A}\over (z-z')} 
\ee

Note that
\be
J^a(z) \psi^{\gamma C}(w)\sim {1\over 2} {1\over z-w} (\sigma^{aT})^\gamma{}_\beta \psi^{\beta C}
\ee

The above OPE algebra gives the commutation relations
\begin{eqnarray}
\com{J^a_m}{J^b_n} &=& \frac{m}{2}\delta^{ab}\delta_{m+n,0} + i{\epsilon^{ab}}_c J^c_{m+n}
            \\
\com{J^a_m}{G^\alpha_{\dot{A},n}} &=& \frac{1}{2}{(\sigma^{aT})^\alpha}_\beta G^\beta_{\dot{A},m+n}
             \\
\ac{G^\alpha_{\dot{A},m}}{G^\beta_{\dot{B},n}} &=& \hspace*{-4pt}\epsilon_{\dot{A}\dot{B}}\bigg[
   (m^2 - \frac{1}{4})\epsilon^{\alpha\beta}\delta_{m+n,0}
  + (m-n){(\sigma^{aT})^\alpha}_\gamma\epsilon^{\gamma\beta}J^a_{m+n}
  + \epsilon^{\alpha\beta} L_{m+n}\bigg]\\
\com{L_m}{L_n} &=& \frac{m(m^2-\frac{1}{4})}{2}\delta_{m+n,0} + (m-n)L_{m+n}\\
\com{L_m}{J^a_n} &=& -n J^a_{m+n}\\
\com{L_m}{G^\alpha_{\dot{A},n}} &=& \left(\frac{m}{2}-n\right)G^\alpha_{\dot{A},m+n}
\end{eqnarray}

\end{document}